\documentclass[twoside]{elsarticle}
\usepackage{array}
\usepackage{units}
\usepackage{multirow}
\usepackage{algorithm2e}
\usepackage{amstext}
\usepackage{graphicx}

\makeatletter

\providecommand{\tabularnewline}{\\}

\usepackage{algorithmicx}
\usepackage[caption=false,font=footnotesize]{subfig}
\usepackage{algpseudocode}
\usepackage{amsfonts}
\usepackage{amsmath}
\usepackage{amssymb}
\usepackage[hang,flushmargin]{footmisc}

\journal{Elsevier Nano Communications Networks}

\LinesNumbered
\SetKwRepeat{Do}{do}{while}%

\newtheorem{theorem}{Theorem}[section]
\newtheorem{lemma}[theorem]{Lemma}

\newenvironment{proof1}[1][Proof.]{\begin{trivlist}
\item[\hskip \labelsep {\bfseries #1}]}{\end{trivlist}}

\@ifundefined{showcaptionsetup}{}{%
 \PassOptionsToPackage{caption=false}{subfig}}
\usepackage{subfig}
\makeatother

\begin{document}

\title{Packet Routing in 3D Nanonetworks: A Lightweight, Linear-path Scheme}

\author[FORTH]{A.~Tsioliaridou}

\ead{atsiolia@ics.forth.gr}

\author[FORTH]{C.~Liaskos}

\ead{cliaskos@ics.forth.gr}

\author[FEMTO-ST]{E.~Dedu\corref{cor1}}

\ead{eugen.dedu@univ-fcomte.fr}

\author[FORTH]{S.~Ioannidis}

\ead{sotiris@ics.forth.gr}

\cortext[cor1]{Corresponding author}

\address[FORTH]{Foundation of Research and Technology-Hellas, Heraklion, Greece}

\address[FEMTO-ST]{Univ. Bourgogne Franche-Comt\'e / FEMTO-ST Institute/CNRS, Montb\'eliard, France}

\begin{abstract}
Packet routing in nanonetworks requires novel approaches, which
can cope with the extreme limitations posed by the nano-scale. Highly
lossy wireless channels, extremely limited hardware capabilities and
non-unique node identifiers are among the restrictions. The present
work offers an addressing and routing solution for static 3D nanonetworks
that find applications in material monitoring and programmatic property
tuning. The addressing process relies on virtual
coordinates from multiple, alternative anchor point sets that act
as \emph{viewports}. Each viewport offers different address granularity
within the network space, and its selection is optimized by a packet
sending node using a novel heuristic. Regarding routing, each node
can deduce whether it is located on the linear segment connecting
the sender to the recipient node. This deduction is made using integer
calculations, node-local information and in a stateless manner, minimizing
the computational and storage overhead of the proposed scheme. Most
importantly, the nodes can regulate the width of the linear path,
thus trading energy efficiency (redundant transmissions) for increased
path diversity. This trait can enable future adaptive routing schemes.
Extensive evaluation via simulations highlights the advantages of
the novel scheme over related approaches.
\end{abstract}
\maketitle

\section{Introduction\label{SECINTRO}}

Nanonetworking is rapidly gaining ground as a key-enabler for novel
industrial and medical applications~\cite{Akyildiz.2010b}. Mobile
nanonetworks are envisioned in collaborating swarms of nano-bots \cite{boillot2015large} or as highly efficient, programmable drug
delivery and virus detection systems within biological organisms.
Static nanonetworks find applications in the monitoring of mission-critical
industrial materials, such as nuclear reactor shielding. However,
constructing and operating a network comprising numerous, nano-sized
nodes creates new technical challenges. This paper focuses on the
networking layer and studies the open research issues of nano-node
addressing and data routing. The novel solution is applicable to large,
static 3D nanonetworks, with omnidirectional antennas and very small node communication radius.
Such networks exchange data packets in a massive multi-hop fashion.
They find exotic applications within active materials, which can receive
external commands and tune their electromagnetic behavior accordingly~\cite{Liaskos.2015}.

Technical and physical limitations at nano-scale call for novel approaches
to nano-node addressing and packet routing~\cite{Wang.2013}. Power
supply units constitute one of the most critical factors under research.
Systems with autonomous nano-nodes must rely on energy scavenging
modules, which presently yield enough power for $1$ packet transmission
per approximately $10$ seconds~\cite{Jornet.2012b}. Wireless power
transfer offers a more effective alternative, at the expense of requiring
an external power supply~\cite{nanocomnet.2015}. Wireless nano-communication
modules, which are expected to operate at the THz band, pose additional
limitations, translating to highly lossy channel conditions due to
acute molecular absorption phenomena~\cite{Wang.2013}. Finally,
manufacturing restrictions and cost scalability considerations correspond
to cheap nano-node hardware, i.e., limited CPU capabilities and data
storage potential~\cite{Akyildiz.2010b}. The impact of these restrictions
on nano-node addressing is that assigning unique identifiers to each
node is not scalable, mainly due to power restrictions. Regarding
data routing, on the other hand, the expectedly frequent transmission
failures require a mechanism to balance path redundancy and energy
consumption, while incurring low computational complexity and memory
overhead~\cite{Wang.2013}.

The proposed solution constitutes a routing and addressing scheme
that complies with the nano-scale restrictions. The node addressing
module relies on virtual coordinates, i.e., geo-locations measured
as node distances from a set of anchor points, collectively referred
to as viewport. Many alternative viewports are considered, and it is
shown that each one offers good address granularity to certain network
areas. Thus, the anchor selection depends on the location of the communicating
nodes. Subsequently, a viewport selection heuristic is proposed, which
enables a sender-node to select the best viewport among available
options. Moreover, a stateless routing scheme is proposed, which
runs on top of the addressing scheme. According to it, a node can
deduce whether it is located on the line connecting the packet sender
to the packet destination. Due to its stateless operation, the scheme does not require inter-node clock synchronization or protocol handshake \cite{d2015timing}. It offers tunable path redundancy
by controlling the width of the 3D line, which constitutes its major
advantage over related solutions~\cite{tsioliaridou15}. Additionally,
it requires node-local information and integer computations only,
respecting the nano-CPU limitations. Finally, due to its stateless
nature, it does not require routing tables for its operation and,
hence, no permanent storage as well.

The remainder of this paper is organized as follows.  Related studies
are given in Section \ref{SECRW}. An overview of nanonode addressing is given in Section \ref{SECADDRESSING}.  Section \ref{SECLINEARROUTING}
details the novel concept and introduces the proposed routing scheme. Evaluation
via simulations takes place in Section \ref{SECSIMULATIONS}. Finally,
the conclusion is given in Section \ref{SECCONCL}.

\section{Related work\label{SECRW}}
The proposed Stateless Linear-path Routing scheme (SLR) is a joint coordinate and routing system. Therefore, in this section we shortly outline the relevant work in both fields and point out the differentiation points of SLR.

\subsection{Localization-Coordinate systems for wireless networks}

The localization and ranging of nodes in wireless networks can be set based on signal indication attributes (such as the received signal strength), on the signal time of flight, or on coordinate systems. The SLR localization and addressing approach falls into the coordinate systems category, with numerous related solutions pertaining to wireless networks~\cite{guvenc2009survey}:

\begin{itemize}
\item Polar coordinate systems, extensively used in radar systems, employing a vector length as one coordinate, as in SLR, but an angle as second coordinate.  Thus, this coordinate system has only one origin, whereas SLR employs three separate origins.
\item Geographic coordinate systems, which are specific to Earth, and allow every location on the Earth to be specified by its latitude, longitude and altitude.  Such coordinates are specific to the surface of Earth, and are thusly different than SLR. Moreover, such systems typically employ more than four origins for improved accuracy, and distances towards them are measured via signal time of flight. The location of the origins is fixed and known in advance. Floating point precision is required to convert between origin distances and the location coordinates. In contrast, in SLR distances are expressed as \emph{integer} numbers, and only integer processing is internally used. Moreover, the locations of the the origins remain unknown to network nodes.
\item Trilinear coordinates, applied in wireless sensor networks, which use three references: the coordinates are the distances to the three sidelines (i.e.\ the infinite line that contain the edge) of a triangle.  In contrast, SLR uses points as origins.
\end{itemize}


SLR can be classified as a particular case of curvilinear system.  Curvilinear coordinate systems are systems in which the coordinate lines ($Ox$, $Oy$ and/or $Oz$) may be curved instead of straight line, cf.\ Fig.~\ref{fig:CurvGranSelecAnch}.
More precisely, SLR is equivalent to the elliptic coordinate system, which is defined as a system in which coordinate lines (i.e.\ the lines obtained when one of the coordinates is kept constant) are confocal ellipses and hyperbolas.  Coordinates $(x,y)$ in SLR system of origins $A_1(0,0,a)$ and $A_2(0,0,-a)$ can be obtained from elliptic coordinates $(u,v)$ as follows:
\begin{equation}
  (x,y) = (a(u+v), a(u-v))
\end{equation}
and reciprocally:
\begin{equation}
  (u,v) = \left(\frac{x+y}{2a}, \frac{x-y}{2a}\right)
\end{equation}

\subsection{Routing}

Related studies on data routing within propagation focused on setups
pertaining to Body Area Network (BAN) applications~\cite{Piro.2015}.
These networks comprise sparse, full-mesh topologies of mobile nodes.
Due to their low numbers, the nodes are commonly assumed to have unique
identifiers. Furthermore, BAN-oriented studies assume hierarchical
networks, where a set of relatively powerful nano-routers control
the smaller, weaker and cheaper nanonodes~\cite{Pierobon.2014b}. In
this context, studies have focused on Medium Access Control (MAC)
schemes that take into consideration the energy harvesting rate to
achieve perpetual operation~\cite{Afsharinejad.2015,mohrehkesh2015drih}.
Subsequently, energy-aware neighborhood discovery and node handshake
processes are proposed in these studies. Due to the presence of the
larger nano-routers, the hierarchical approach is too intrusive for
the smart material applications targeted by the present study\ \cite{Liaskos.2015}.
Additionally, such applications require large networks, and minimal
transmission power per node, which translates to multi-hop routing
requirements and non-unique addressing.

Due to their mutual emphasis on low-overhead communication, routing
in nanonetworks exhibits similarities to wireless sensor networks
(WSN)~\cite{YU.2015}. However, the restrictions of the nano-environment
invalidate existing WSN routing solutions. It has been experimentally
confirmed that well-known routing protocols, such as AODV, DSDV, or
DSR are not scalable in terms of nodes or energy expenditure\ \cite{lemmon2009geographic}.
More lightweight geo-addressing WSN approaches provide satisfactory
performance for large networks~\cite{lemmon2009geographic}. Nonetheless,
such routing algorithms require directional routing, attained with
neighborhood discovery, routing tables, and even memorization of past
node attributes (last-seen location, direction, velocity).

Similarities also exist between routing
in nanonetworks and networks-on-chips (NoCs)~\cite{Younis.2014}.
NoCs need to discover their topology and perform defect mapping in
a completely decentralized manner~\cite{Catania.2014}. Nonetheless,
NoCs assume few, powerful nodes, which rely on external power supply
and not on scavenging~\cite{Younis.2014}. Therefore, NoC-oriented
solutions are generally not portable to nanonetworks.

Regarding the authors' prior work, a ray-tracing-based simulation
technique for nanonetworks was studied in~\cite{Kantelis.2014b}.
Liaskos et al.\ proposed a metaheuristic-based, selective flooding
dissemination scheme for 2D nanonetworks~\cite{LiaskosTC.2014},
which was later refined in terms of complexity~\cite{nanocomnet.2015}.
Tsioliaridou et al.\ proposed a joint peer-to-peer routing and geo-addressing
approach for 2D networks~\cite{tsioliaridou15}, which was subsequently
extended to 3D~\cite{Tsioliaridou1n3}. However, the routing module
in these studies operated via selective flooding within network \emph{volumes}.
The present paper differentiates by proposing a routing approach over
curvilinear paths, accomplished in a stateless manner.

An early version of this work, which does not include the viewport
selection optimization process, can be found in~\cite{Tsioliaridou2015SLR}.

\section{Nanonode addressing in 3D\label{SECADDRESSING}}

\begin{figure}
\centering\includegraphics[width=0.6\columnwidth]{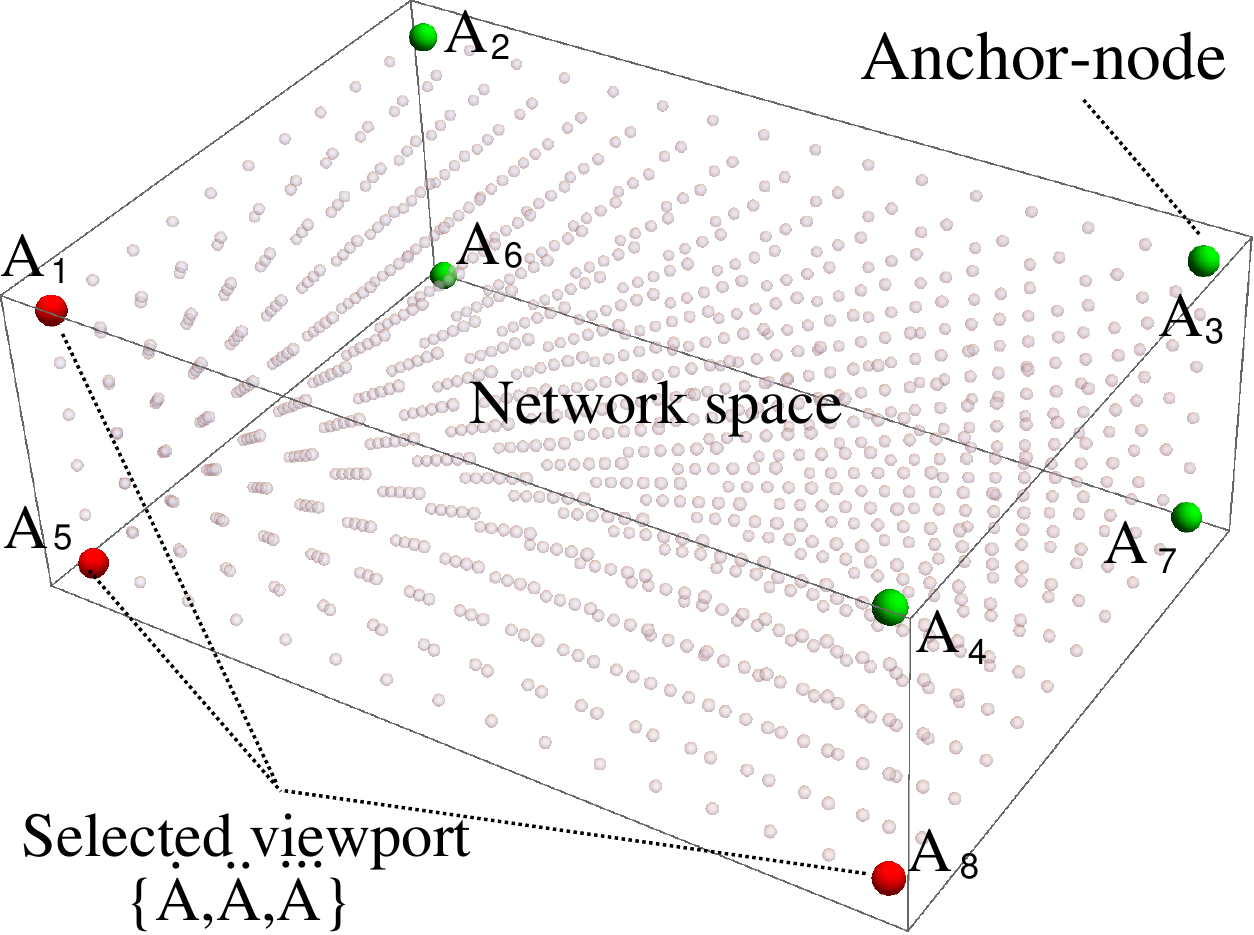}
\caption{\label{fig:Overview-of-the}Overview of the studied nanonetwork.}
\end{figure}

Figure~\ref{fig:Overview-of-the} illustrates the considered system
setup, comprising a set of nanonodes placed within a 3D rectangular
\emph{network space}. The layout of the nodes is either a regular
grid or random within the space. The grid setup corresponds to an
active meta-material application, while the random placement to a
smart material monitoring setup~\cite{Liaskos.2015}. The network
nodes have the same hardware.  Protocols using mixed omni- and uni-directional antennas have been proposed~\cite{yao16}.
The presented SLR scheme assumes uniformity: each node is equipped with identical omnidirectional antennas.
This is in line with its restricted-flood-based operating principle:
no next node has to be explicitly chosen during the SLR packet routing process.
As such, they have a short wireless connectivity radius
imposed by the nature of the application~\cite{Liaskos.2015}, and
we assume a multi-hop packet routing case. Additionally, the network
conditions are such that node failures are common, accentuating the
need for alternative paths (path redundancy). Nanonode communication
may temporarily fail due to error-prone hardware, energy shortage
or channel conditions~\cite{Jornet.2012}.

Eight nodes, denoted as \emph{anchors}, are placed at the vertexes of
the space during the construction of network. These anchors are indexed
as shown in Fig.~\ref{fig:Overview-of-the} ($A_{1\ldots8}$). Note
that the anchors are identical to any other node. Their uniqueness
pertains to their role in the node \emph{addressing} phase.

\textbf{Node addressing.} This stage happens once and serves as
initialization of the system. It allocates addresses to nanonodes,
which are saved for the lifetime of the network. We employ the 3D
location of a node as its address using a trilateration process~\cite{trilateration}.
Initially, the nodes obtain their distances from the anchors as follows.
The anchor $A_{1}$ initiates the addressing phase by broadcasting
a data packet with the structure shown in Table~\ref{tab:Structure-of-setup}.

\begin{table}
\centering\begin{tabular}{|c|c|c|}
\hline
setup flag (1 bit) & anchor index (3 bits) & hop count\tabularnewline
\hline
\end{tabular}
\caption{\label{tab:Structure-of-setup}Structure of setup packets.}
\end{table}

A \emph{setup flag} set to the value ``1'' denotes that the packet
is exchanged as part of the initialization phase. Additionally, the
\emph{hop count} integer field is set to $0$. Each non-anchor recipient
node increases the hop count field by $+1$ and memorizes the ensuing
value as its distance from anchor $A_{anchor\,index}$. Subsequently,
it re-broadcasts the packet. These steps are executed only for the
first received packet from a given anchor. Should an anchor-node with
index $\left(anchor\,index\right)+1$ receive such a packet (e.g.,
$A_{2}$), it is triggered to generate its own setup packet. A trivial
timeout is allowed to ensure that the completion of the previous setup
packet has been completed. Thus, once the initialization phase is
complete, each node has obtained its address as hop distances,
$\left\{ r_{i},\,i=1\ldots8\right\}$, from the anchors $A_{1\ldots8}$.

Notice that several neighboring nodes are assigned the same address.
In other words, an address refers to an area, denoted as \emph{zone},
rather than a node. The fact that all nodes within a zone share the
same geo-address allows for a degree of natural node fail-over. Given
the frequent node failures in the nano-environment, geo-address sharing
increases the chances that the network will remain connected after
each failure.

If for any reason a node does not receive at the right time the setup packet sent by an anchor, it can still receive it via the retransmissions of its neighboring nodes. Thus, the node will be still connected, albeit mapped at slightly different coordinates (off by~1 hop).  Having some nodes mapped at slightly different positions could add a few spurious communications, without, however, posing a problem to the protocol.

Generally, a point in 3D space is uniquely identified by its distances
from $4$ anchor points~\cite{trilateration}. However, the employed
3D addressing scheme can conditionally use only 3 anchors for this
task. The necessary condition for unique zone identification follows.

\begin{lemma}
\label{lemma1}
A zone can be uniquely identified by three anchors (\emph{viewport})
distances, $\left\{ \dot{r},\,\ddot{r},\,\dddot{r}\right\}$, located at
the same face of the rectangular network space.
\end{lemma}

\begin{proof1}
Assume a rectangular network space with side lengths $X,\ Y,\ Z$, and a
Cartesian system, with $x\in\left[0,\ X\right]$, $y\in\left[0,\ Y\right]$,
$z\in\left[0,\ Z\right]$. Let a viewport comprising three anchors located
on one face of the space:
\begin{equation}
\dot{A}=\left\{ 0,\ 0,\ 0\right\} ,\,\ddot{A}=\left\{ X,\ 0,\ 0\right\} ,\,\dddot{A}=\left\{ 0,\ 0,\ Z\right\}
\end{equation}
Notice that, by proper rotation and transfer, these anchors can represent
any viewport on the same side of a 3D rectangle. Furthermore, let
any point $P$ within the space have Cartesian coordinates, $\left\{ x,\ y,\ z\right\}$,
and distances from the anchors, $\left\{ \dot{r},\,\ddot{r},\,\dddot{r}\right\}$.
These distances fulfil the following equations:
\begin{equation}
z=\sqrt{\dot{r}^{2}-x^{2}-y^{2}}\label{eq:r}
\end{equation}
\begin{equation}
z=\sqrt{\ddot{r}^{2}-\left(x-X\right)^{2}-y^{2}}\label{eq:rt}
\end{equation}
\begin{equation}
z=Z-\sqrt{\dddot{r}^{2}-x^{2}-y^{2}}\label{eq:rtt}
\end{equation}
We will show that each $\left\{ \dot{r},\,\ddot{r},\,\dddot{r}\right\}$
corresponds to a unique $\left\{ x,\ y,\ z\right\}$ triplet. From
(\ref{eq:r}) and (\ref{eq:rt}) we easily derive $x$ uniquely as:
\begin{equation}
x=\frac{r^{2}-\ddot{r}{}^{2}+X^{2}}{2\cdot X}\label{eq:floatx}
\end{equation}
$z$ is also uniquely identified from (\ref{eq:r}) and (\ref{eq:rtt})
as:
\begin{equation}
z=\frac{\dot{r}^{2}-\text{\ensuremath{\dddot{r}}}^{2}+Z^{2}}{2\cdot Z}\label{eq:floaty}
\end{equation}
Finally, from (\ref{eq:r}) we obtain two candidate $y$ values as:
\begin{equation}
y=\pm\sqrt{\dot{r}^{2}-x^{2}-z^{2}}\label{eq:floatz}
\end{equation}
However, our initial assumption is that $y\in\left[0,\ Y\right]$.
Thus, the negative solution can be rejected, leading to the unique
definition of the complete triplet, $\left\{ x,\ y,\ z\right\}$.
$\blacksquare$
\end{proof1}

The proven property enables the linear routing scheme detailed in
the ensuing Section.

\section{Stateless Linear Routing\label{SECLINEARROUTING}}

The presented Stateless Linear Routing (SLR) defines a way for routing a packet
from any sender node $P_1$ to any recipient $P_2$, both identified by
their addresses. The sender initially selects one viewport out of
$A_{1\ldots8}$ (i.e., anchors $\left\{ \dot{A},\ddot{A},\dddot{A}\right\}$), which is essentially the coordinate system\emph{ (CS)} for the packet
route towards node $P_{2}$. The considerations on the viewport selection
process are discussed in the dedicated subsection below. The
corresponding \emph{usable addresses (UA)} of $P_1$ and $P_2$ are denoted
as $UA_{1}:\left\{ \dot{r}_{1},\ddot{r}_{1},\dddot{r}_{1}\right\}$ and
$UA_{2}:\left\{ \dot{r}_{2},\ddot{r}_{2},\dddot{r}_{2}\right\}$.  Then,
the sender constructs a packet structured as shown in
Table~\ref{tab:Packet-format,-as}, where \emph{setup flag} is~0, \emph{packet id} is a random
integer number and \emph{DATA} is the useful load of the packet. Any node
$P$ receiving this packet: i) checks if a packet with the same id has been
already received, at which case it discards the packet. Elsewhere, the node:
ii) memorizes the packet id for a trivial amount of time, in order to
avoid route loops. iii) Deduces its usable address, $UA:\left\{
\dot{r},\ddot{r},\dddot{r}\right\}$, based on the CS field. iv)
Re-transmits the packet if it is located on the linear path connecting
$P_{1}$ and $P_{2}$. The latter check is performed as follows.

\begin{table}
\centering\begin{tabular}{|c|c|c|}
\hline
setup flag (1 bit) & packet id (8 bits) & CS (3$\times3$ bits)\tabularnewline
\hline
UA$_{1}$ (var) & UA$_{2}$ (var) & DATA (var)\tabularnewline
\hline
\end{tabular}
\caption{\label{tab:Packet-format,-as}Packet format, as constructed by the
sender.}
\end{table}

In order to deduce whether a node $P$ is located on the straight
line, $\overline{P_{1}P_{2}}$, one should ideally convert the UAs
of $P,\,P_{1},\,P_{2}$ to Cartesian coordinates, and then check the
compliance to the standard relation:
\begin{equation}
\frac{x-x_{1}}{x_{2}-x_{1}}=\frac{y-y_{1}}{y_{2}-y_{1}}=\frac{z-z_{1}}{z_{2}-z_{1}}
\end{equation}
However, this approach poses two challenges. First, converting UAs
to Cartesian coordinates requires knowledge of the overall network
space dimensions and floating point processing capabilities, as shown
in equations (\ref{eq:floatx})--(\ref{eq:floatz}). Notice that the
nodes are only aware of their distances from the network anchors.
Deriving the network space dimensions requires extra computations
with floating point precision as well.

In order to avoid the complexity challenges, we shall treat the UAs,
i.e., point-to-anchor distances $P:\left\{ \dot{r},\ddot{r},\dddot{r}\right\}$,
as a curvilinear coordinate system~\cite{curvilinear}. Thus, the
equation of a line connecting two points $P_{1}:\left\{ \dot{r}_{1},\ddot{r}_{1},\dddot{r}_{1}\right\}$
and $P_{2}:\left\{ \dot{r}_{2},\ddot{r}_{2},\dddot{r}_{2}\right\}$
becomes:
\begin{equation}
\frac{\dot{r}-\dot{r}_{1}}{\dot{r}_{2}-\dot{r}_{1}}=\frac{\ddot{r}-\ddot{r}_{1}}{\ddot{r}_{2}-\ddot{r}_{1}}=\frac{\dddot{r}-\dddot{r}_{1}}{\dddot{r}_{2}-\dddot{r}_{1}}\label{eq:linearDistances}
\end{equation}
This approach also justifies the focus of Lemma~\ref{lemma1} to
three anchors, instead of the four required by formal trilateration
processes.

The second challenge for linearity check stems from the fact that
the distances $\dot{r},\ddot{r},\dddot{r}$ measure number of hops
and, therefore, are integer numbers. Thus, relation (\ref{eq:linearDistances})
will not hold precisely in the general case. To address this issue
we rewrite relation (\ref{eq:linearDistances}) as:
\begin{equation}
\left\{ \begin{array}{c}
\left(\dot{r}-\dot{r}_{1}\right)\left(\ddot{r}_{2}-\ddot{r}_{1}\right)-\left(\ddot{r}-\ddot{r}_{1}\right)\left(\dot{r}_{2}-\dot{r}_{1}\right)=0\\
\left(\dot{r}-\dot{r}_{1}\right)\left(\dddot{r}_{2}-\dddot{r}_{1}\right)-\left(\dddot{r}-\dddot{r}_{1}\right)\left(\dot{r}_{2}-\dot{r}_{1}\right)=0
\end{array}\right.\label{eq:alt1}
\end{equation}
Subsequently, we define the quantities:
\begin{equation}
\mathbf{\Delta^{a}}\left(\dot{r},\ddot{r}\right)=\left(\dot{r}-\dot{r}_{1}\right)\left(\ddot{r}_{2}-\ddot{r}_{1}\right)-\left(\ddot{r}-\ddot{r}_{1}\right)\left(\dot{r}_{2}-\dot{r}_{1}\right)\label{eq:deltaA}
\end{equation}
\begin{equation}
\mathbf{\Delta^{b}}\left(\dot{r},\dddot{r}\right)=\left(\dot{r}-\dot{r}_{1}\right)\left(\dddot{r}_{2}-\dddot{r}_{1}\right)-\left(\dddot{r}-\dddot{r}_{1}\right)\left(\dot{r}_{2}-\dot{r}_{1}\right)\label{eq:deltaB}
\end{equation}
Therefore, instead of evaluating direct compliance with (\ref{eq:linearDistances})
or (\ref{eq:alt1}), we can simply check whether $\mathbf{\Delta^{a}}$,
$\mathbf{\Delta^{b}}$ undergo a sign change when altering $\dot{r},\,\ddot{r},\,\dddot{r}$.
Thus, we define the quantities:
\begin{equation}
\begin{array}{cc}
\mathbf{\Delta_{\dot{r}}^{a}}\left(\dot{r},\ddot{r}\right)=\mathbf{\Delta^{a}}\left(\dot{r}-m,\ddot{r}\right)\\
\mathbf{\Delta_{\ddot{r}}^{a}}\left(\dot{r},\ddot{r}\right)=\mathbf{\Delta^{a}}\left(\dot{r},\ddot{r}-m\right)\\
\mathbf{\Delta_{\dot{r}\ddot{r}}^{a}}\left(\dot{r},\ddot{r}\right)=\mathbf{\Delta^{a}}\left(\dot{r}-m,\ddot{r}-m\right)
\end{array}
\end{equation}
\begin{equation}
\begin{array}{cc}
\mathbf{\Delta_{\dot{r}}^{b}}\left(\dot{r},\dddot{r}\right)=\mathbf{\Delta^{b}}\left(\dot{r}-m,\dddot{r}\right)\\
\mathbf{\Delta_{\dddot{r}}^{b}}\left(\dot{r},\dddot{r}\right)=\mathbf{\Delta^{b}}\left(\dot{r},\dddot{r}-m\right)\\
\mathbf{\Delta_{\dot{r}\dddot{r}}^{b}}\left(\dot{r},\dddot{r}\right)=\mathbf{\Delta^{b}}\left(\dot{r}-m,\dddot{r}-m\right)
\end{array}
\end{equation}
where $m$ is a non-zero integer. Thus, a node can deduce whether
it is placed on a line by performing the following checks:
\begin{equation}
\begin{array}{c}
\left[\mathbf{\Delta^{a}}\cdot\mathbf{\Delta_{\dot{r}}^{a}}\le0
\quad\mbox{OR}\quad
\mathbf{\Delta^{a}}\cdot\mathbf{\Delta_{\ddot{r}}^{a}}\le0
\quad\mbox{OR}\quad
\mathbf{\Delta^{a}}\cdot\mathbf{\Delta_{\dot{r}\ddot{r}}^{a}}\le0\right]\\
\mbox{AND}\\
\left[\mathbf{\Delta^{b}}\cdot\mathbf{\Delta_{\dot{r}}^{b}}\le0
\quad\mbox{OR}\quad
\mathbf{\Delta^{b}}\cdot\mathbf{\Delta_{\dddot{r}}^{b}}\le0
\quad\mbox{OR}\quad
\mathbf{\Delta^{b}}\cdot\mathbf{\Delta_{\dot{r}\dddot{r}}^{b}}\le0\right]
\end{array}\label{eq:tre}
\end{equation}
The integer $m$ controls the ``width'' of the linear path, essentially
introducing a way of meeting the requirement for path redundancy in
nanonetworking. A value of $m=1$ corresponds to minimal redundancy,
while greater values increase the number of alternative paths. The
value of $m$ could be a global preset, or a part of the packet header
set by the original sender. For example, a sender may initially choose
a value of $m=1$. If no delivery acknowledgment is received, the
sender can then retry with a value of $m=2$, etc. The redundancy
introduced by $m$ is complementary to the zone redundancy, enforced
by the addressing scheme as discussed in Section~\ref{SECADDRESSING}.
Zone redundancy favors network connectivity, while $m$ introduces
path redundancy by employing additional zones.

The condition (\ref{eq:tre}) checks whether a node is located on
a linear path of width $m$, defined by $P_1$ and
$P_2$. Additionally, a node can check whether it is on the line
\emph{segment} connecting $P_1$ to $P_2$ based on the condition:
\begin{multline}
\left(\dot{r}-\dot{r}_{2}\right)\left(\dot{r}-\dot{r}_{1}\right)\le0
\quad\mbox{AND}\quad
\left(\ddot{r}-\ddot{r}_{2}\right)\left(\ddot{r}-\ddot{r}_{1}\right)\le0
\quad\mbox{AND}\quad\\
\left(\dddot{r}-\dddot{r}_{2}\right)\left(\dddot{r}-\dddot{r}_{1}\right)\le0\label{eq:seg}
\end{multline}
Therefore, the linear routing process, outlined as Algorithm~\ref{alg:SLR},
consists of checking compliance with condition~(\ref{eq:seg})
first, and subsequently with condition~(\ref{eq:tre}). If any condition
yields false, the packet is consumed and is not retransmitted.

\begin{algorithm}
\caption{\label{alg:SLR}The Stateless Linear Routing process.}
\hrule
\KwIn{An incoming packet $\texttt{pkt}$ at a node $P$.}
\KwOut{Retransmission decision.}
\hrule
Derive $UA:\left\{ \dot{r},\ddot{r},\dddot{r}\right\}$ of $P$
from $\texttt{pkt.CS}$

$UA_{1}\gets\texttt{pkt.UA}_{1}$, $UA_{2}\gets\texttt{pkt.UA}_{2}$

$m\gets\texttt{pkt.m}$~~~~~~~~~~~~~~~~// in case
$\texttt{m}$ is a header of $\texttt{pkt}$

\If{Condition(15) $\texttt{AND}$ Condition(16)}{retransmit $\texttt{pkt}$}
\hrule
\end{algorithm}

Note that the described packet routing process was designed to meet
the nanonetworking specifications outlined in Section~\ref{SECINTRO}.
It inherently provides tunable path redundancy via the $m$ parameter,
which also affects the involved number of retransmitters and, hence,
the expended energy. Additionally, its stateless nature (absence of
routing tables) translates to minimal memory overhead. Moreover, it
considers the limited nano-CPU capabilities, by requiring integer
calculations only.

\subsection{Viewport optimization\label{subsec:Viewport-optimization}}

\begin{figure}
\centering\includegraphics[width=0.55\columnwidth]{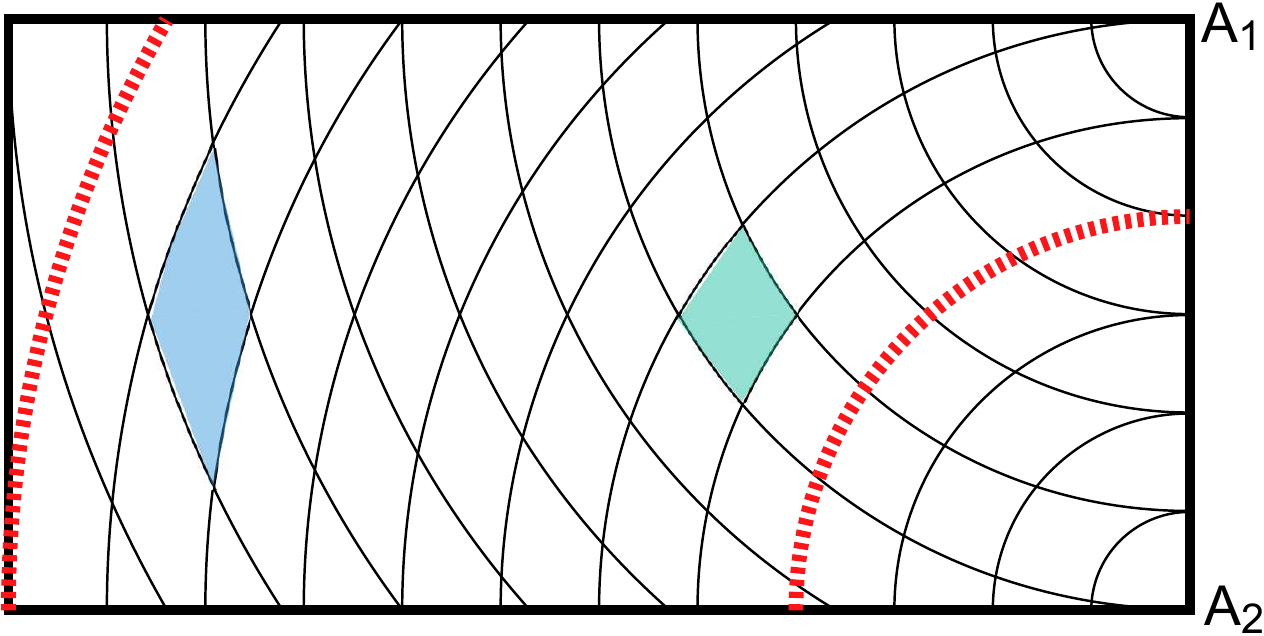}
\caption{\label{fig:CurvGranSelecAnch}The path curvature and zone resolution
trade-off stemming from the selection of anchors.}
\end{figure}

A consideration stemming from the viewport selection process is illustrated
in the 2D example of Fig.~\ref{fig:CurvGranSelecAnch}. The direct
mapping of curvilinear coordinates to Cartesian naturally yields curved
paths (dashed lines). Should we select anchors $A_{1}$,~$A_{2}$
too close to a communicating pair the curvature is high, while distant
anchors yield approximately straight paths. On the other hand, the
zone resolution of the addressing system is better (i.e., fewer nodes
per zone) for medium pair-to-anchors distances (shaded areas, Fig.~\ref{fig:CurvGranSelecAnch}).
Ideally, the path should be straight, with high zone-granularity.
We proceed to study this aspect in the current subsection.

According to the described routing process, the packet origin node,
$P_{1}$, is given the option to select the optimal viewport (coordinate
system comprising an anchor triplet) that will handle the complete
packet propagation to $P_{2}$. According to Lemma~\ref{lemma1},
there exist $24$ valid viewport options ($4$ on each face of the
space), denoted as the set~$\mathcal{CS}$. Furthermore, each selection
affects the involved number of retransmitters, as shown qualitatively
in Fig.~\ref{fig:CurvGranSelecAnch}. We will assume that the number
of retransmitters per volume unit is constant within the network space.
Therefore, we seek the viewport, $cs_{o}\in\mathcal{CS}$, that minimizes
the volume defined by condition~(\ref{eq:tre}) $\texttt{AND}$ (\ref{eq:seg})
as follows:

\begin{equation}
cs_{o}=\text{argmin}_{(cs\in\mathcal{CS})}\left\{ \int_{x=0}^{X}\int_{y=0}^{Y}\int_{z=0}^{Z}Predicate\left(\left(\ref{eq:tre}\right)\texttt{\,AND\,}\left(\ref{eq:seg}\right)\right)\,dx\,dy\,dz\right\} \label{eq:optimizeViewFormal}
\end{equation}
where $Predicate(*)$ is a function that returns $1$ when condition
$*$ is true, and $0$ otherwise. Additionally, $\left\{ x,y,z\right\}$
are the Cartesian coordinates corresponding to a triplet $\left\{ \dot{r},\ddot{r},\dddot{r}\right\}$
via equations (\ref{eq:floatx}), (\ref{eq:floaty}) and (\ref{eq:floatz}),
while the triplet $\left\{ \dot{r},\ddot{r},\dddot{r}\right\}$ itself
is a function on the chosen viewport, $cs$.

The exact calculation of equation (\ref{eq:optimizeViewFormal}) is
prohibitive for a hardware-constrained nano-node, which is also assumed
capable for integer-processing only in our case. Thus, we propose an alternative,
heuristic approach that complies with these computational restrictions.
The heuristic logic relies on model-fitting: instead of solving (\ref{eq:optimizeViewFormal})
to obtain $cs_{o}$, we search for the $cs$ that most closely matches
a good model-viewpoint.
\begin{figure}
\centering\includegraphics[bb=0bp 0bp 600bp 300bp,clip,width=.9\columnwidth]{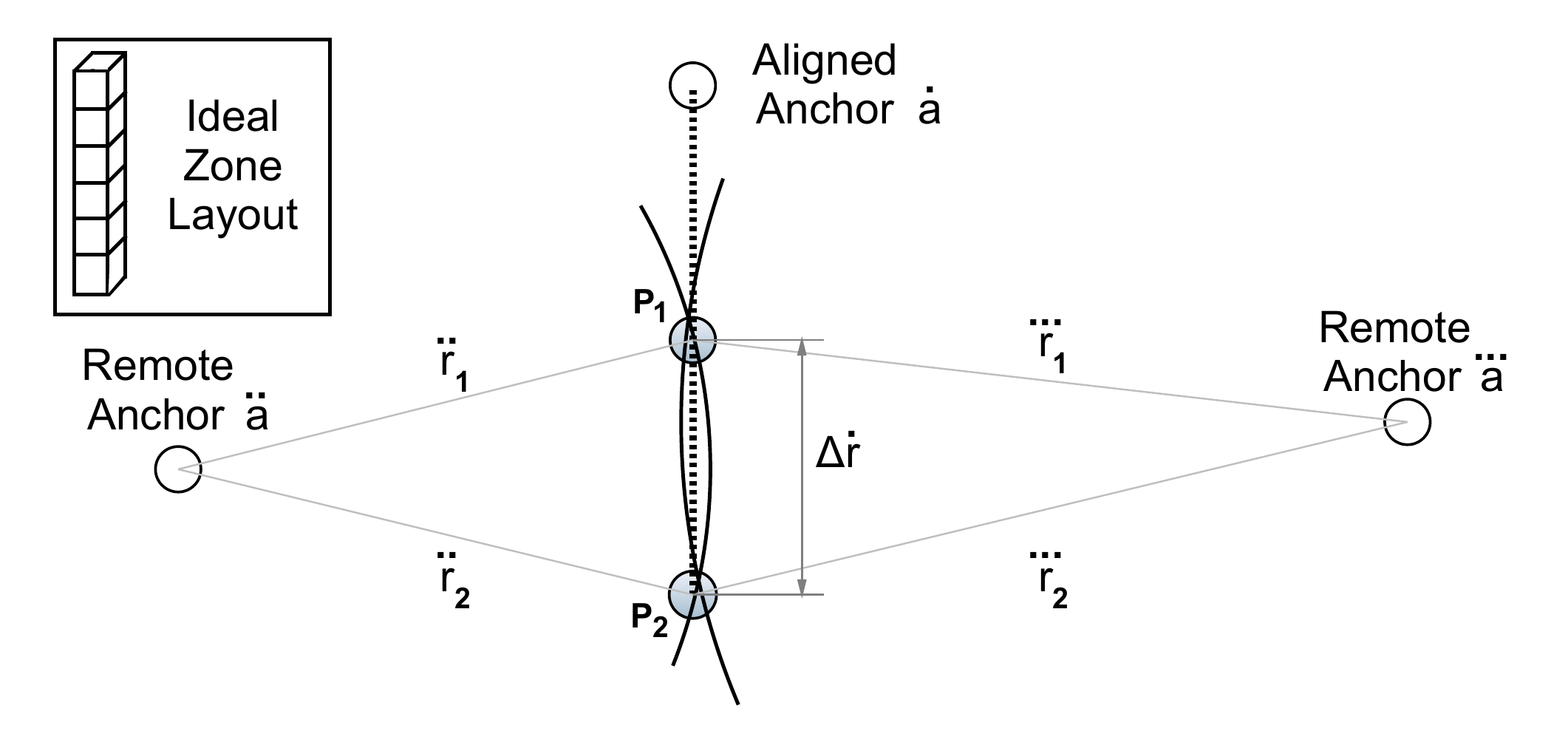}
\caption{\label{fig:ViewpointModel}The model-viewpoint for a communicating
node pair $P_{1},\,P_{2}$.}
\end{figure}

The rationale for the employed model-viewpoint is illustrated in Fig.~\ref{fig:ViewpointModel}.
The model comprises three anchors, in accordance with Lemma~\ref{lemma1},
which are denoted as one \emph{aligned anchor} $\dot{\text{a}}$,
and two \emph{remote anchors} $\ddot{\text{a}}$, $\dddot{\text{a}}$.
The aligned anchor is located on the line defined by nodes $P_{1}$
and $P_{2}$. In this manner, a maximal zone resolution over the straight
path connecting $P_{1}$ to $P_{2}$ is ensured, as per the ideal
layout shown in Fig.~\ref{fig:ViewpointModel}. In contrast, an anchor,
e.g., perpendicular to the segment $\overline{P_{1}P_{2}}$ would
see limited or no zone differentiation over it. Remote anchors $\ddot{\text{a}}$,
$\dddot{\text{a}}$ are located as far as possible from nodes $P_{1}$
and $P_{2}$. Ideally, this translates to negligible curvature over
nodes $P_{1}$ and $P_{2}$, producing an almost straight path. The
aligned and remote anchors combined yield the ideal zone layout of
Fig.~\ref{fig:ViewpointModel}.

Finding the anchor that best matches the aligned anchor is straightforward.
Essentially, we seek the anchor for which the quantity $\Delta\dot{r}$
in Fig.~\ref{fig:ViewpointModel} is maximized. Let $A_{i}$, $i=1\ldots8$
be the set of all anchors of the network space. Moreover, let $R\left(n_{1},n_{2}\right)$
be the distance between two nodes $n_{1}$ and $n_{2}$. Then, the
aligned anchor $\dot{\text{a}}$ is best approximated as:

\begin{equation}
\dot{\text{a}}\gets argmax{}_{(A_{i})}\left\{ \left|R\left(P_{1},A_{i}\right)-R\left(P_{2},A_{i}\right)\right|\right\} \label{eq:bestAlignedAnchor}
\end{equation}
Finding the farthest anchor from $P_{1}$ \emph{and} $P_{2}$
(either $\ddot{\text{a}}$ or $\dddot{\text{a}}$) is accomplished
as follows:

\begin{equation}
\ddot{\text{a}}\gets argmax{}_{(A_{i})}\left\{ \min\left\{ R\left(P_{1},A_{i}\right),\,R\left(P_{2},A_{i}\right)\right\} \right\} \label{eq:mostRemoteAnchor}
\end{equation}
In other words, the distance of an anchor from a \emph{pair} of nodes
$P_{1}$, $P_{2}$ as a whole, is defined as their minimum distance
from the anchor. Notice that both relations (\ref{eq:bestAlignedAnchor})
and (\ref{eq:mostRemoteAnchor}) uphold the integer computations-only
restriction.

Having defined the aligned and remote anchor approximations, we proceed
to define a procedure to obtain the viewpoint that best fits the model
of Fig.~\ref{fig:ViewpointModel}. The process requires a prioritization
of either the aligned or the remote anchor definition. For instance,
selecting the aligned anchor $\dot{\text{a}}$ first via relation
(\ref{eq:bestAlignedAnchor}) means that the possible viewport choices
will now be limited to those containing the chosen $\dot{\text{a}}$
only (i.e., $9$ viewports out of $24$). A similar narrowing of
choices occurs with the prioritization of $\ddot{\text{a}}$. The
prioritization essentially refers to either good zone resolution (by
choosing $\dot{\text{a}}$), or a shorter path between $P_{1}$ and
$P_{2}$ but with larger zones. With no loss of generality, we prioritize
good zone resolution since, as shown in Fig.~\ref{fig:CurvGranSelecAnch},
zones can be enlarged and deformed significantly at great distances
from the anchors. Thus, the achieved zone layout would differ significantly
from the ideal one given in Fig.~\ref{fig:ViewpointModel}.

The viewport selection process, run by the sender $P_{1}$ when preparing
to send an original packet, is formulated as Algorithm~\ref{alg:VPSEL}.
At step $1$, the process selects the aligned anchor in accordance
with relation (\ref{eq:bestAlignedAnchor}). Steps 2, 3 initialize
two helper variables that will be utilized to update and eventually
hold the best viewport. Steps 4--11 focus on all viewport choices
that contain the selected aligned anchor. Each candidate viewport
$cs$ is broken down to its contents in step $5$, while the ordering
of $\ddot{\text{a}}$ and $\dddot{\text{a}}$ is arbitrary. In order
to reduce complexity by avoiding a double sorting by $\ddot{\text{a}}\to\left(P_{1}P_{2}\right)$
and $\dddot{\text{a}}\to\left(P_{1}P_{2}\right)$ distances (relation
(\ref{eq:mostRemoteAnchor})), step $6$ calculates a collective distance
of anchor-pair to node-pair, $\left(\ddot{\text{a}},\dddot{\text{a}}\right)\to\left(P_{1}P_{2}\right)$,
using the $min()$ reduction as well. Steps 7--10 update the
helper variables, holding the $cs$ that contains the most aligned
\emph{and} the most remote anchors. Finally, the usable addresses
of $P_{1}$ and $P_{2}$ are set at steps 12--13 and are incorporated
to the packet format of Table~\ref{tab:Packet-format,-as}.

\begin{algorithm}
\caption{\label{alg:VPSEL}The viewport selection process.}

\hrule
\KwIn{The distances of nodes $P_{1}$, $P_{2}$ from the anchors
$A_{1\ldots8}$.}

\KwOut{The usable addresses $UA_{1}$ $UA_{2}$ of $P_{1}$, $P_{2}$.}
\hrule
$\dot{\text{a}}\gets argmax{}_{(A_{i})}\left\{ \left|R\left(P_{1},A_{i}\right)-R\left(P_{2},A_{i}\right)\right|\right\}$

$best\_cs\gets\emptyset$

$max\_distance\gets-1$

\For{ $cs\,\in\mathcal{CS}:$~$\dot{\text{a}}\in cs$} {

$\left\{ \ddot{\text{a}}\right.$,$\left.\dddot{\text{a}}\right\} \gets$$cs-\dot{\text{a}}$

$D\gets \min\left\{ R\left(P_{1},\ddot{\text{a}}\right),\,R\left(P_{2},\ddot{\text{a}}\right),\,R\left(P_{1},\dddot{\text{a}}\right),\,R\left(P_{2},\dddot{\text{a}}\right)\right\}$

\If{$D>max\_distance$}{
$best\_cs\gets cs$

$max\_distance\gets D$
}
}

$UA_{1}\gets P_{1}(best\_cs)$

$UA_{2}\gets P_{2}(best\_cs)$
\hrule
\end{algorithm}

Regarding its complexity, Algorithm~\ref{alg:VPSEL} requires:
\begin{itemize}
\item 16 operations for step $1$ ($2$ operations for each of the $8$
anchors),
\item 2 operations for steps 2--3 (collectively),
\item 81 operations for the for loop ($9$ operations for each of the
$9$ iterations),
\item 2 operations for steps 12--13 (collectively),
\end{itemize}
leading to a total of 101 integer calculations. Its memory requirements
are 6 integer slots required to hold the variables $\dot{\text{a}}$,
$\ddot{\text{a}}$, $\dddot{\text{a}}$, $best\_cs$, $max\_distance$,
$D$.

\section{Simulations\label{SECSIMULATIONS}}

This Section evaluates the routing efficiency of the proposed Stateless
Linear Routing (SLR), compared to the CORONA routing approach~\cite{tsioliaridou15}.
CORONA is a recently proposed, stateless and provenly scalable nano-routing
approach. It proposes a packet flood approach, contained within the
area defined by the minimal and maximal anchor distances of the communicating
node-pair. Thus, although originally proposed for 2D networks, CORONA
has easily been extended for our purposes to the 3D case as well. The simulator is implemented
on the AnyLogic platform~\cite{XJTechnologies.2013}.

Given that CORONA does not include a viewpoint optimization approach,
the evaluation is broken into two parts. Firstly, we evaluate the
proposed, model-based viewpoint selection for SLR only (subsection~\ref{subsec:evalVP}).
Algorithm~\ref{alg:VPSEL} is compared to: i) the optimal viewpoint
selection derived via brute force, and ii) random viewpoint selection.
Versions prioritizing zone resolution and anchor distance are studied.
The metric of viewpoint selection performance is the number of involved
retransmitters per communicating node-pair (fewer is better). Secondly,
we evaluate the communication efficiency of SLR versus CORONA (subsection~\ref{subsec:evalNE}),
keeping a static viewpoint for the duration of the evaluation. The
main comparison metric is the communication success probability between
any node pair, under the conditions of: i) failures of intermediate
nodes due to energy depletion, and ii) simultaneous, potentially interfering
communications taking place in parallel.

\subsection{Configuration}

The employed simulation configuration is given in Table~\ref{tab:Persistent-Simulation-Parameters}~\cite{nanocomnet.2015}.
We assume a number of $N=5000$ nodes within a cubic 3D network space
with side size 1\,cm. The nodes are placed according to two layouts: i) on a regular grid with $\nicefrac{1}{16}\,\mathrm{cm}$ spacing, or ii) randomly within the space. The latter case is
derived from the regular grid by deactivating a random subset of the
nodes, as discussed below (run setup).

\begin{table}
\centering
\begin{tabular}{|c|c||c||c||c||c|}
\hline
Parameter & \multicolumn{5}{c|}{Value}\tabularnewline
\hline
\hline
\multicolumn{6}{|c|}{\textbf{Communication \& Power Parameters}}\tabularnewline
\hline
Frequency & \multicolumn{5}{c|}{$100$ GHz}\tabularnewline
\hline
Transmission Power & \multicolumn{5}{c|}{$5$ dBnW}\tabularnewline
\hline
Noise Level & \multicolumn{5}{c|}{$0$ dBnW}\tabularnewline
\hline
Reception $\text{SINR}$ threshold & \multicolumn{5}{c|}{$-10$ dB}\tabularnewline
\hline
Guard Interval & \multicolumn{5}{c|}{$0.1$ nsec}\tabularnewline
\hline
Packet Duration & \multicolumn{5}{c|}{$10$ nsec}\tabularnewline
\hline
\multicolumn{6}{|c|}{\textbf{Path Attenuation Parameters}}\tabularnewline
\hline
Absorption Coefficient $K$ & \multicolumn{5}{c|}{$0.52$~dB/Km }\tabularnewline
\hline
Shadow Fading Coefficient $X$ & \multicolumn{5}{c|}{$0.5$~dB}\tabularnewline
\hline
\end{tabular}\smallskip{}
\begin{tabular}{|c|c|}
\hline
\multicolumn{2}{|c|}{\textbf{Space setup (Number of nodes: $5000$)}}\tabularnewline
\hline
\hline
Cubic $1\times1\times1$ cm, & \multirow{2}{*}{\includegraphics[height=0.5cm]{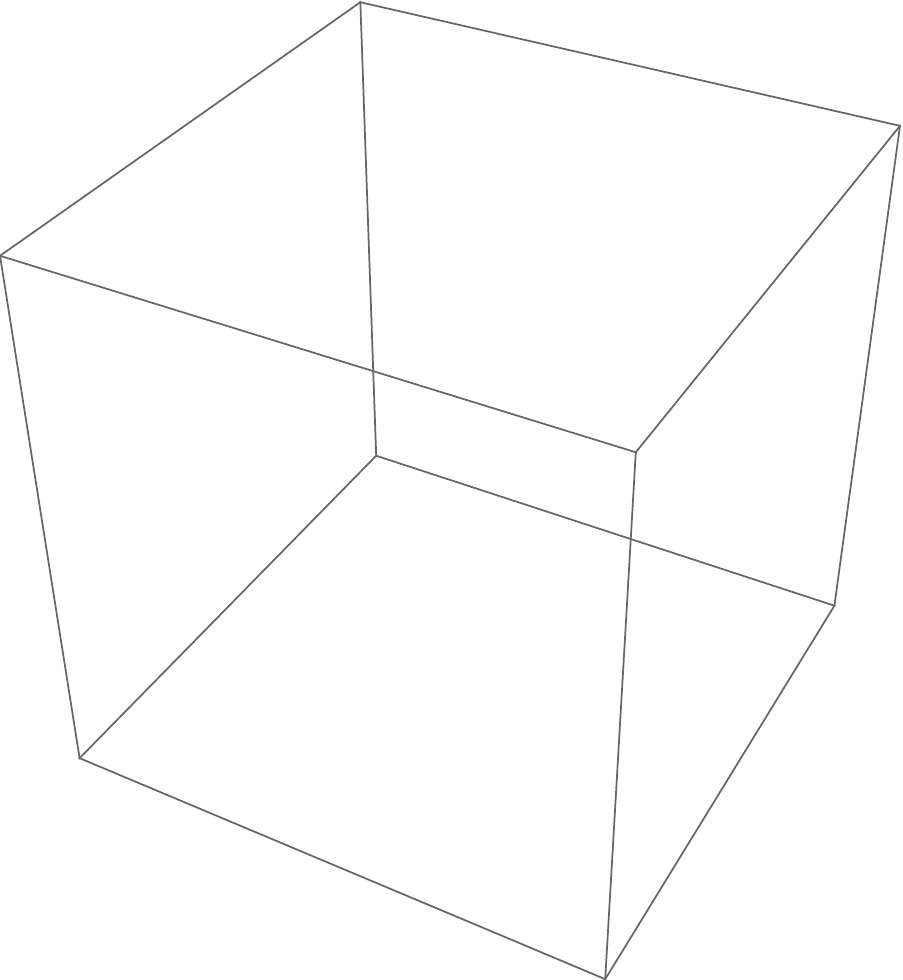}}\tabularnewline
Grid/Random node placement & \tabularnewline
\hline
\end{tabular}
\caption{\label{tab:Persistent-Simulation-Parameters}Simulation Configuration.}
\end{table}

Each node represents a nano-controller in a smart meta-material application~\cite{Liaskos.2015}.
The space among the nodes is filled with air composed of standard
atmospheric gases at normal humidity~\cite{ITUR.Feb.2007}. This
dense network setup as a whole is intended to approximate the building
block for a visible light-interacting smart meta-material~\cite{Liaskos.2015}.

Regarding the communication model, all nodes are equipped with isotropic
antennas. Molecular absorption due to the air (absorption coefficient~$K$~\cite{Jornet.2011})
and shadow fading ($X$ coefficient in~dB~\cite{Kim.2014}) are
taken into account. The Signal to Interference plus Noise (SINR) model
is used to deduce the success of a packet reception~\cite{Iyer.2009,Lehtomaki.2014}.
The transmission power was chosen to ensure that each zone comprises
approximately $15$ nodes. Thus, the resolution of the network space
is $7\times7\times7$ zones within a $1\,\mathrm{cm}^3$ volume. This low
resolution constitutes a \emph{worst-case} scenario for the proposed
scheme, since it hinders the formation of well-defined linear paths.
The behavior in higher resolution cases is also studied at the end
of this Section.

\subsection{Evaluation of the viewpoint selection process\label{subsec:evalVP}}

\begin{figure}
\centering

\subfloat[\label{fig:allapproachesm1}Model-based (Algorithm~\ref{alg:VPSEL})
and random viewpoint selection efficiency (path redundancy, $m$=1).
Both zone resolution and anchor distance prioritizations are examined
for the model-based approach.]
{\includegraphics[width=1\textwidth]{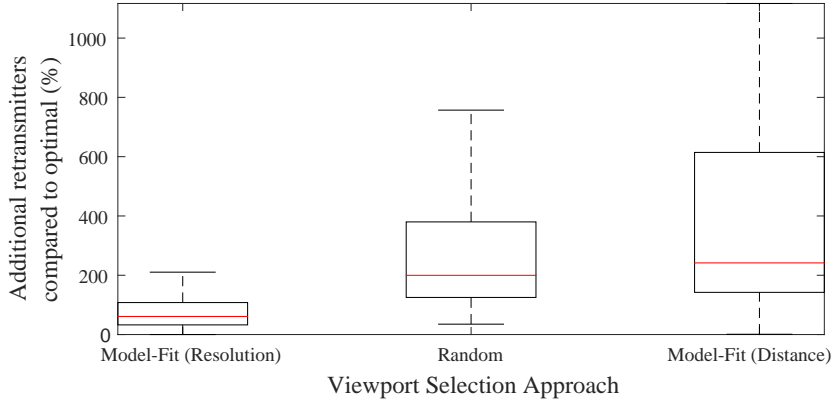}}

\subfloat[\label{fig:Zone_m}Model-based viewpoint selection efficiency (zone
resolution prioritization) for increasing path redundancy, $m$.]
{\includegraphics[width=1\textwidth]{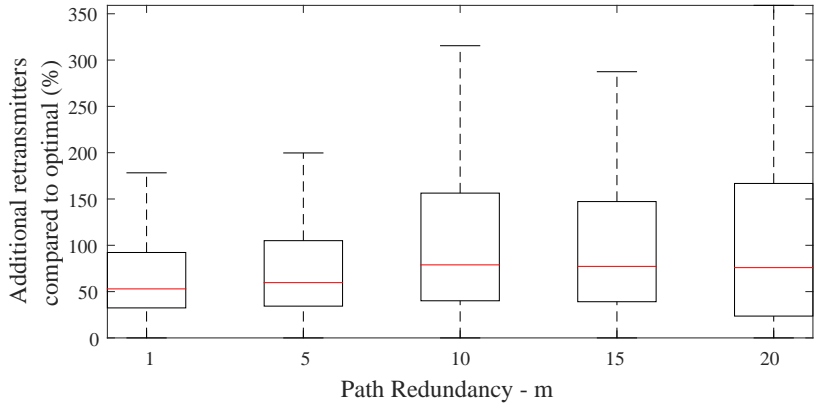}}

\subfloat[\label{fig:random_m}Random viewpoint selection efficiency for increasing path redundancy, $m$.]
{\includegraphics[width=1\textwidth]{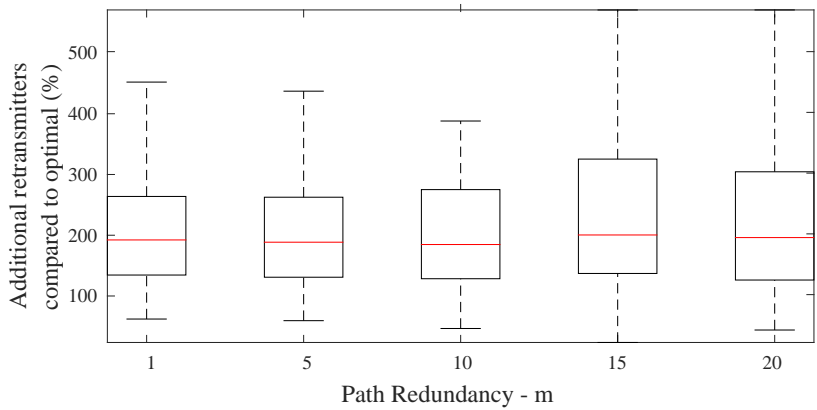}}
\caption{\label{fig:anchorMetrics}Evaluation of viewpoint selection approaches.}
\end{figure}

\textbf{Setup.} The simulation runs are executed as follows. Once
the nodes are placed in their 3D layout (grid or random), we proceed
to randomly select $100$ node pairs in a sequential manner. For each
pair, we execute a packet exchange and log the number of retransmitters
yielded by the proposed viewpoint selection approach presented in
Section~\ref{subsec:Viewport-optimization}. Additionally, we study
the effects of prioritization (zone \emph{resolution} or anchor \emph{distance})
by swapping lines 1 and 6 in Algorithm~\ref{alg:VPSEL}. Moreover,
we exhaustively check all valid coordinate systems ($24$) per pair,
and deduce the one yielding the lowest number of transmitters in each
case (\emph{optimal}). The average number of retransmitters over all
possible viewpoint choices represents the \emph{random} viewport selection
approach. For clarity, all results are relative to the optimal performance.
All nodes are considered powered-on in this experiment. The results
over all pairs are gathered in the from of boxplots, presented in
Fig.~\ref{fig:anchorMetrics}. The results are similar for both random
and grid node layouts.

\textbf{Results.} Initially, in Fig.~\ref{fig:allapproachesm1},
it is observed that the model-based selection approach with resolution
prioritization outperforms all alternatives. The \emph{resolution}
approach yields $50\%$ more retransmitters than the optimal case
on average, which is encouraging for a lightweight scheme that is
restricted to employ integer processing only. Note that random selection
yields $200\%$ more retransmitters than the optimal case, which is
indicative of the gains achieved by the proposed approach. Moreover,
the prioritization of zone resolution is shown to be superior to anchor
distance, as intuitively expected in Section~\ref{subsec:Viewport-optimization},
i.e., the distance prioritization produces large zones, leading to
a performance similar or worse than the random selection.

We proceed to study the behavior of model-based (resolution prioritization)
and random viewport selection as the path redundancy parameter increases.
The rationale for this check is that linear paths get thicker as $m$
increases, meaning that the effects of improper viewpoint selection
could be magnified. Model-fitting with distance prioritization is
not considered due to the preceding results of Fig.~\ref{fig:allapproachesm1}.
As it can be seen in Fig.~\ref{fig:Zone_m} (model-based with zone resolution)
and Fig.~\ref{fig:random_m} (random), both viewpoint selection approaches
yield a good degree of independence from the path redundancy parameter
$m$. The model-based approach retains a performance of 50--75\%
additional retransmitters than the optimal case, while the random
selection is nearly flat around $200\%$. This phenomenon has a natural
explanation. The $m$ parameter is itself a magnifying factor of the
volume of retransmitters. The greater the $m$ values, the (proportionately)
thicker the linear path. However, a viewport represents an orthogonal
concern, i.e., the curvature of the linear path, regardless of its
thickness. Thus, the viewpoint selection yields a good degree of independence
from the $m$ parameter.

\subsection{Evaluation of the networking efficiency\label{subsec:evalNE}}

\textbf{Setup.} A run is initiated by forming the described grid arrangement
of nodes. A random percentage of nodes (denoted as \emph{deactivation
ratio}) is deactivated, emulating failing nodes. Thus, non-zero values
of the deactivation ratio have the indirect effect of randomizing
the topology as well. Then, a series of $100$ operation cycles takes
place, with an interarrival of $10$ seconds as follows. A number of nodes
pairs (denoted as \emph{pair number}) are randomly selected,
each requiring the exchange of a single, unique packet. At the end
of the $100$ cycles, we log the percentage of pairs that communicated
successfully (\emph{comm. success ratio}). Each run is
repeated $100$ times, randomizing the node failures anew, to improve
the confidence of the presented results in various topologies.

\begin{figure}
\centering\includegraphics[width=.7\columnwidth]{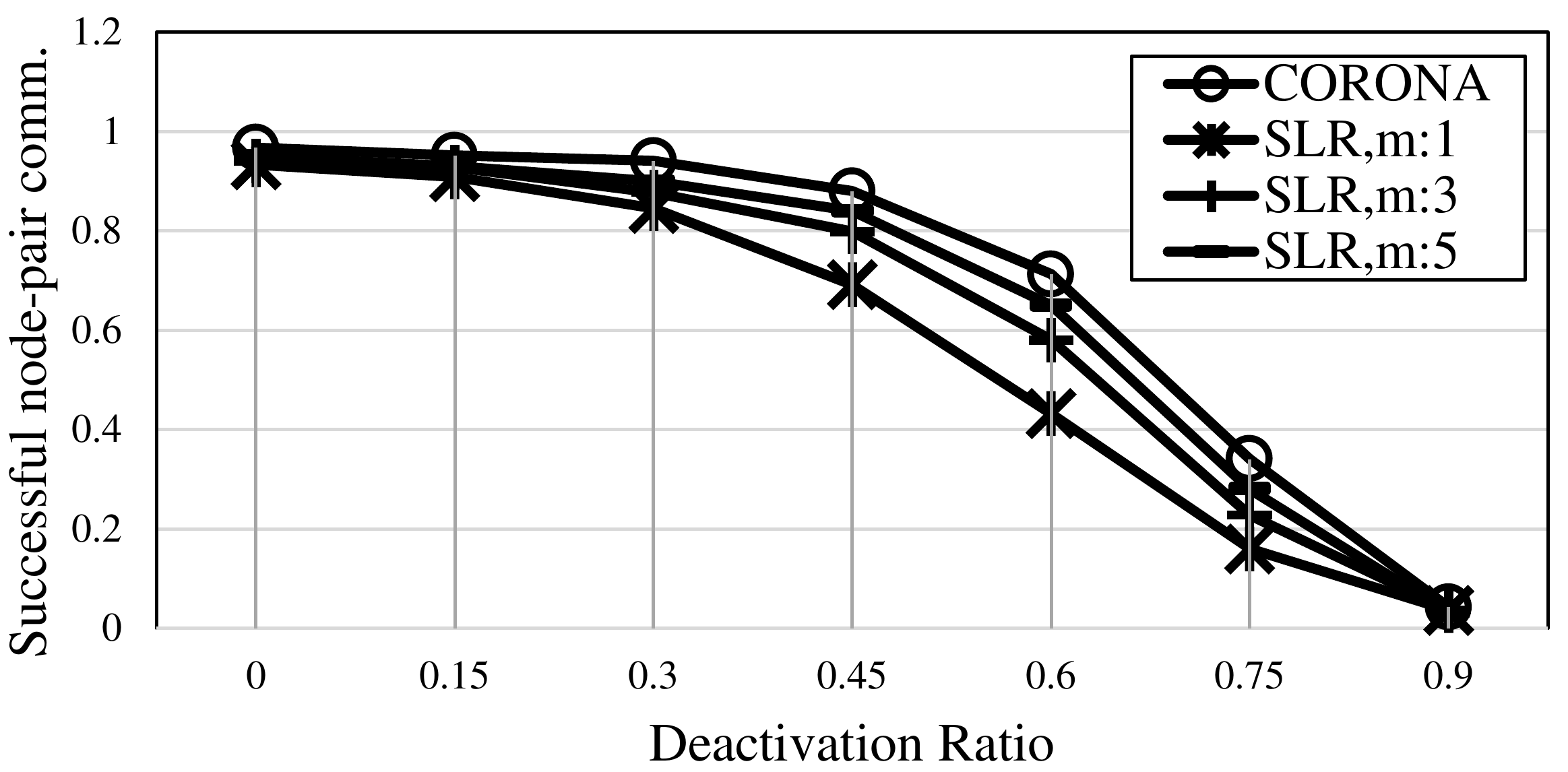}
\caption{\label{fig:SUCC}SLR tunability effects on the node-pair communication
ratio.}
\end{figure}

\begin{figure}
\centering\includegraphics[width=.7\columnwidth]{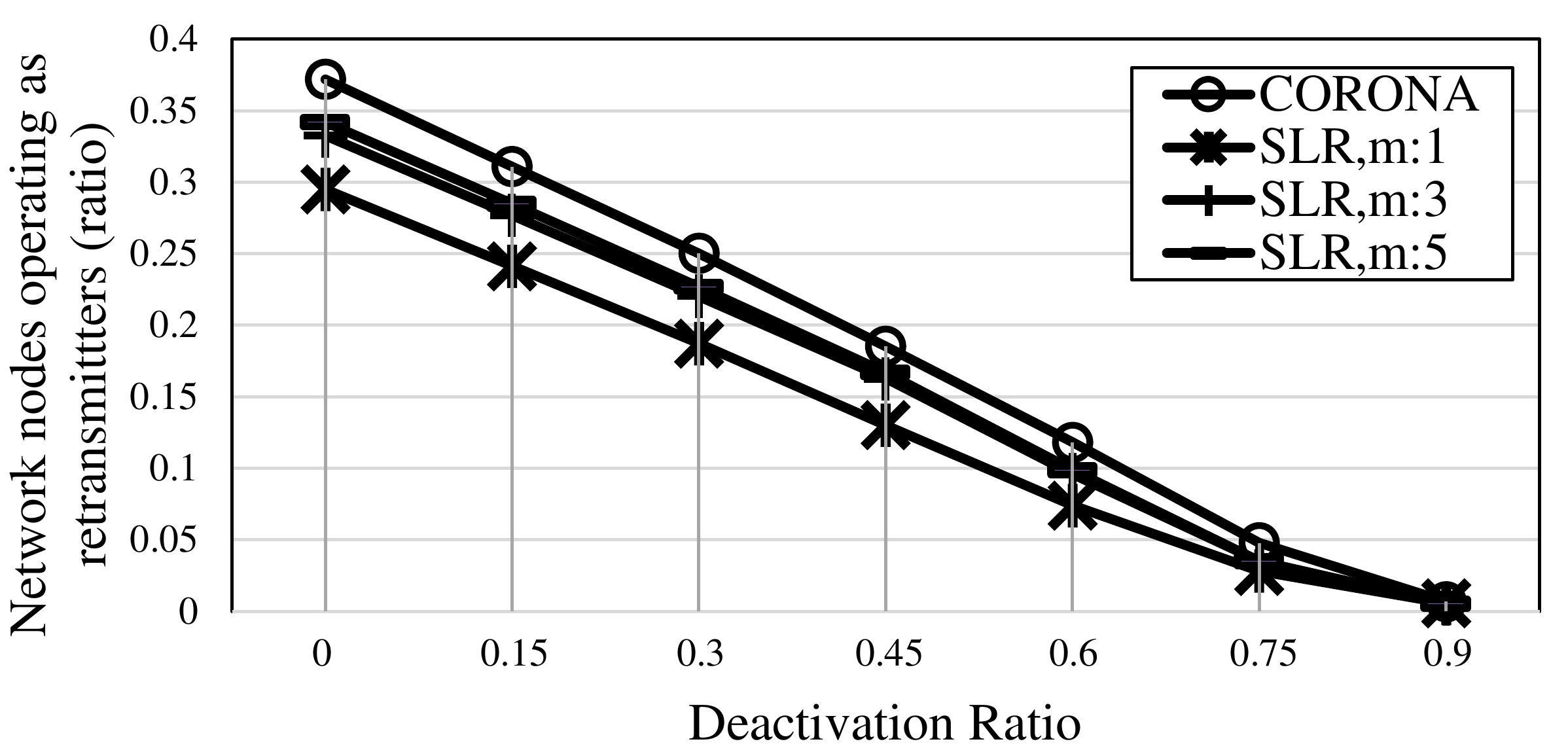}
\caption{\label{fig:RETRANS}SLR tunability effects on the average, network-wide
ratio of retransmitters serving each communicating pair.}
\end{figure}

\textbf{Results.} Figures \ref{fig:SUCC} and \ref{fig:RETRANS}
illustrate the tunability of the proposed scheme (SLR) versus the
related CORONA approach. The pair number is kept constant to $5$,
and the deactivation ratio varies from $0$ to $90\%$. Figure\ \ref{fig:SUCC}
then shows the attained communication success ratio for various level
of SLR path redundancy ($m$), while Figure\ \ref{fig:RETRANS}
presents the corresponding average percentage of the total network
nodes serving as intermediate retransmitters per communicating pair.
A higher number of retransmitters is evidence of higher energy expenditure
rate. The plots in both Figures naturally decrease as the deactivation
ratio increases. High deactivation rates yield segmented network
areas, reducing the success ratio. Additionally, fewer nodes remain
available to serve as retransmitters.

Notably, SLR introduces tunable success ratio and energy efficiency
via the path redundancy parameter $m$, as described in the analysis
of Section\ \ref{SECLINEARROUTING}. In other words, SLR allows for
a sender node to regulate the network energy expenditure rate, depending
on its estimation for the network state. Thus, future schemes can
exploit SLR to automatically employ less path redundancy (and, thus,
less expense energy) when the network conditions are estimated as
relatively good. On the opposite case, when the network conditions
are characterized as challenging, the sender node could automatically
increase $m$, attaining higher path diversity and communication success
ratio. Mechanisms for network state estimation and automatic adaptation
constitute possible extensions of the present work.

\begin{figure}
\centering\includegraphics[width=.7\columnwidth]{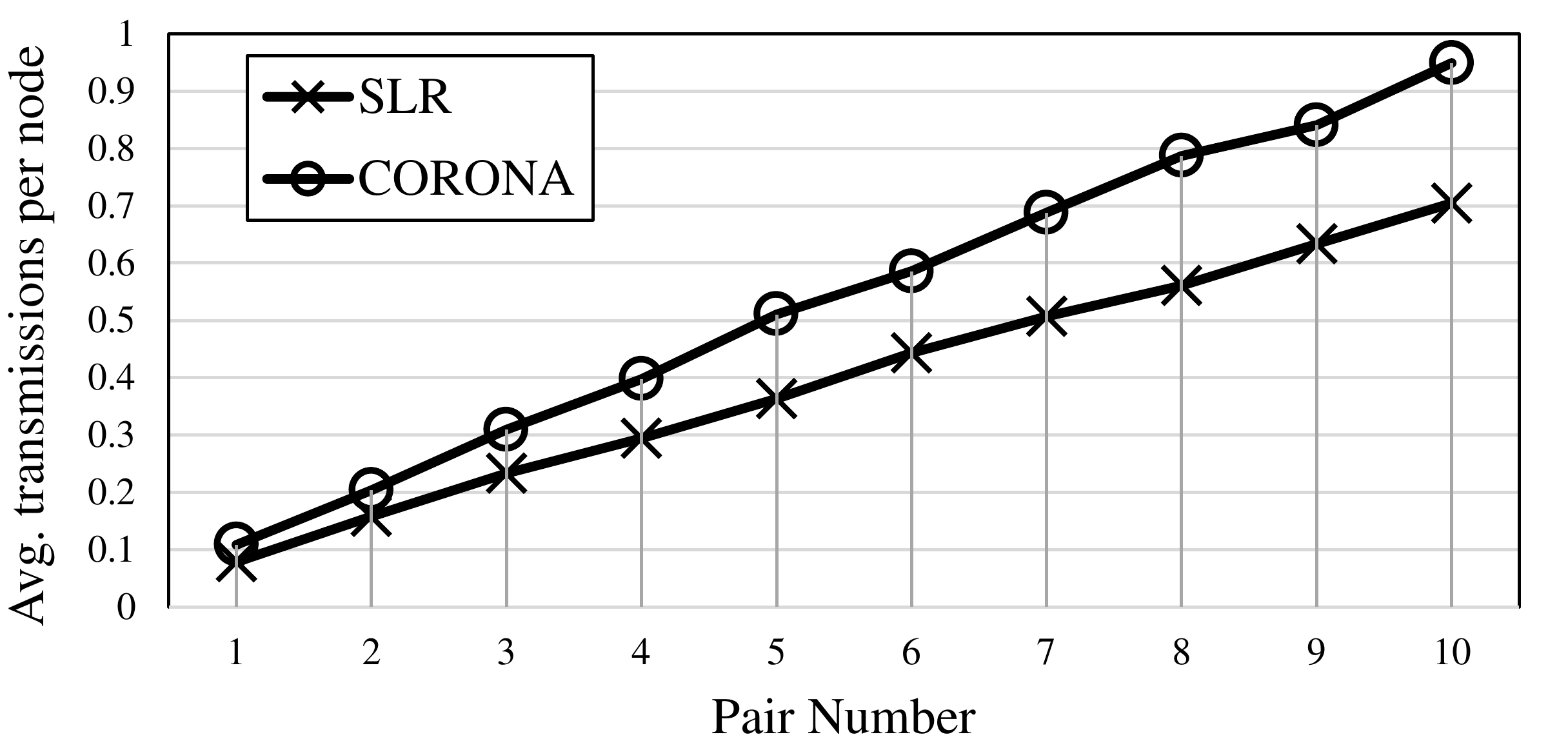}
\caption{\label{fig:PARPAIRS}Average packet transmissions per node imposed
by CORONA and the proposed SLR, versus the number of concurrently
communicating node pairs in the network.}
\end{figure}

We proceed to evaluate the potential for parallel communications offered
by SLR in Fig.\ \ref{fig:PARPAIRS}. Keeping the path redundancy
and deactivation ratio constant ($m=1$ and $0\%$ respectively),
the pair number is varied in the range 1 to 10, i.e., one to
ten node pairs communicating in parallel (x-axis). For each case,
we measure the average number of retransmissions imposed to each network
node. A higher number of retransmissions implies that the network
nodes will deplete their energy reserved faster, thus yielding lower
potential for parallel communications. SLR provides better performance
over CORONA from this point of view, which is attributed to the well-defined,
linear form of the SLR paths. On the other hand, CORONA defines much
larger volumes within which packet flooding is executed. Thus, it inevitably
incurs more redundant transmissions per node, yielding decreased parallel
communications potential.

\begin{figure}
\centering\includegraphics[width=0.5\columnwidth]{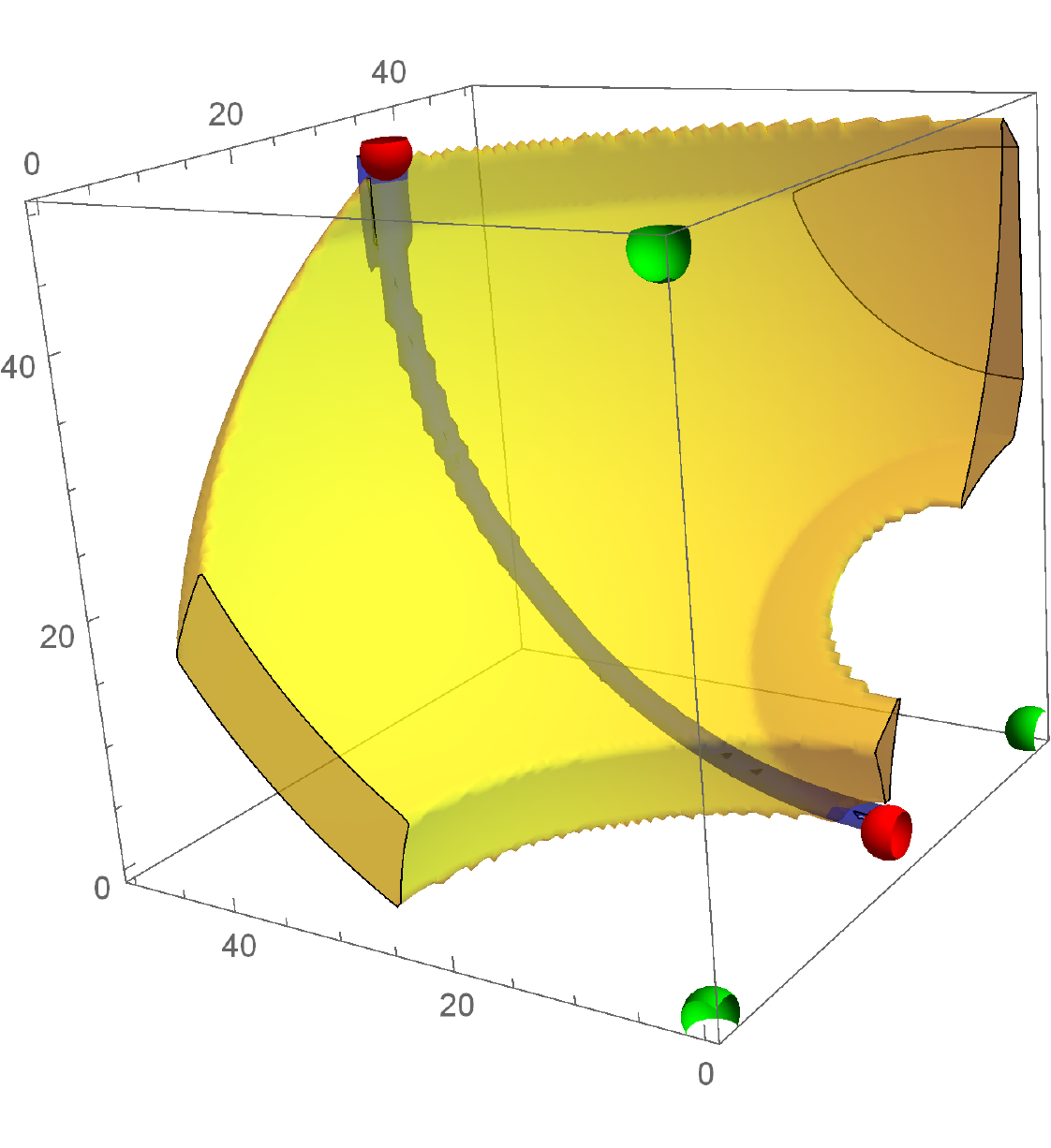}
\caption{\label{fig:CORVSLIN}Comparison of CORONA (yellow area) and the novel
SLR (dark curve) routing behavior for a given communicating node pair
in a \emph{high-resolution} space.}
\end{figure}

The presented results are especially promising, given that they refer
to a very small space ($1\,\mathrm{cm}^3$), where the gain margin is very
limited due to the low zone resolution (cf.\ ``Configuration'' above).
In Fig.\ \ref{fig:CORVSLIN} we proceed to study the expected gains
in a bigger network space with dimensions $50\times50\times50\,\mathrm{cm}$.
The number of nodes increases proportionally and yields simulation
runtime issues, therefore we restrict Fig.\ \ref{fig:CORVSLIN}
to a single, indicative node-pair communication case ($m=1$). As
illustrated, the gains of SLR over CORONA increase considerably in
bigger spaces. SLR yields a well-defined arc (i.e., curvilinear path).
On the other hand, CORONA produces a much larger volume of restransmitters,
defined by the distances of the communicating nodes from the selected
anchors. Thus, CORONA trades routing area size for wider network reachability
(i.e., high path redundancy), albeit without tunability potential.
SLR, on the other hand, can yield very narrow, linear paths to very
wide network areas, setting the basis for adaptivity in the highly
challenging nano-environment.

\section{Conclusion and future work\label{SECCONCL}}

The present study introduced a novel addressing and routing scheme
for 3D electromagnetic nanonetworks. The scheme allows for tunable
routing path redundancy, in order to counter the highly lossy nature
of nano-communications while limiting redundant transmissions. Additionally,
it yields well-defined, linear routing paths among communicating pairs,
allowing for a considerable degree of parallel transmissions within
the network. The scheme is stateless, requiring no permanent memory
overhead or neighborhood discovery at the nanonode-side. Limitations
of nano-CPUs are also taken into account, and each routing decision
requires few integer calculations only. The traits of the novel scheme
were evaluated via extensive simulations.

Future work includes devising methods to estimate network state in order to adapt line width $m$ automatically,
and compose multiple linear path segments to reach non-line-of-sight targets.

\section*{Acknowledgement}
This work was funded by the European Union via the Horizon 2020: Future Emerging Topics call (FETOPEN), grant EU736876, project VISORSURF (http://www.visorsurf.eu).

\section*{References}

\bibliographystyle{elsarticle-harv}

\end{document}